\documentclass[a4paper,11pt]{article}
\usepackage [latin1]{inputenc}
\usepackage[T1]{fontenc}
\usepackage{default}
\usepackage{color}
\usepackage{amssymb}
\usepackage{amsmath}
\usepackage{graphicx}
\usepackage{mathtools}
\usepackage{ragged2e}

\usepackage{dcolumn}
\usepackage{bm}
\usepackage{graphicx}
\usepackage{braket}
\usepackage{verbatim} 
\usepackage[english]{babel}
\usepackage{hyperref}
\hypersetup{
citecolor=red,
colorlinks=true,
filecolor=red,
linkcolor=blue,
linktocpage=true,
urlcolor=blue
} 
\usepackage[titletoc,toc,title]{appendix}
\numberwithin{equation}{section}
\usepackage{cleveref}
\usepackage[margin=2.5cm]{geometry}
\usepackage{cite}
\usepackage{epsfig}
\usepackage{float}
\usepackage{cancel}
\usepackage{amsfonts}
\usepackage{enumitem}
\usepackage[font={footnotesize,it}]{caption}
\usepackage{authblk}

{\makeatletter \g@addto@macro\bfseries{\boldmath} \makeatother}

\begin{document}

\title{Entanglement negativity in Galilean conformal field theories}

\author[1]{Vinay Malvimat\thanks{\noindent E-mail:~ {\tt vinaymmp@gmail.com}}}
\author[2]{Himanshu Parihar\thanks{\noindent E-mail:~ {\tt himansp@iitk.ac.in}}}
\author[2]{Boudhayan Paul\thanks{\noindent E-mail:~ {\tt paul@iitk.ac.in}}}
\author[2]{Gautam Sengupta\thanks{\noindent E-mail:~ {\tt sengupta@iitk.ac.in}}}

\affil[1]{
Indian Institute of Science Education and Research\\

Homi Bhabha Rd, Pashan, Pune 411 008, India
\bigskip
}

\affil[2]{
Department of Physics\\

Indian Institute of Technology\\ 

Kanpur 208 016, India
}

\date{}

\maketitle

\thispagestyle{empty}

\begin{abstract}

\noindent

\justify

We obtain the entanglement negativity for various bipartite zero and finite temperature pure and mixed state configurations in a class of $(1+1)$-dimensional Galilean conformal field theories. In this context we establish a construction for computing the entanglement negativity for such bipartite states involving a suitable replica technique. Our construction exactly reproduces certain universal features observed for entanglement negativity of corresponding states in relativistic $(1+1)$-dimensional conformal field theories.

\end{abstract}

\clearpage

\tableofcontents

\clearpage

\section{Introduction}
\label{sec1}

\justify

Characterization of quantum entanglement has emerged as a central theme in the study of diverse phenomena ranging from condensed matter physics to issues of quantum gravity and black holes. A significant role in these developments involves the measure of entanglement entropy which is defined by  the von Neumann entropy of the reduced density matrix for a given subsystem of a bipartite quantum system and characterizes quantum entanglement in a pure state. This is simple
to compute for finite systems but often intractable for extended quantum many body systems like quantum field theories. A formal definition of the entanglement entropy for such extended systems may be obtained through the {\it replica technique} but is in general difficult to compute. Interestingly as demonstrated in \cite{Calabrese:2004eu,Entanglement,Calabrese:2005in}, this quantity may be computed in $(1+1)$-dimensional relativistic conformal field theories ($CFT_{1+1}$)
through the above mentioned replica technique.

It is well known however in quantum information theory, that the entanglement entropy
fails to be a viable measure for the characterization of mixed state entanglement as it receives
contributions from correlations irrelevant to the entanglement of the mixed state under consideration. This renders the characterization of mixed state entanglement to be a subtle and complex issue. This issue was addressed in a classic work by Vidal and Werner \cite{Vidal} in which they introduced a computable measure termed {\it entanglement negativity} which characterized the upper bound on the {\it distillable entanglement} of the mixed state under consideration.\footnote{Here the distillable entanglement refers to the total number of Bell states that may be extracted from the mixed state under consideration using only local operations and classical communication (LOCC).} The entanglement negativity was defined as the logarithm of the trace norm for the partially transposed density matrix for a bipartite system with respect to one of the subsystems. It was demonstrated by Plenio in \cite{PhysRevLett.95.090503} that unlike usual entanglement measures the entanglement negativity is non convex, however it has been shown in quantum information theory that it is an entanglement monotone. Interestingly in a series of communications Calabrese et al. in \cite{PhysRevLett.109.130502,Negativity,Calabrese:2014yza} advanced a replica technique to compute the entanglement negativity for bipartite systems described by a $CFT_{1+1}$.

In a separate context a class of $(1+1)$-dimensional Galilean invariant conformal field theories 
$(GCFT_{1+1})$ has been obtained in \cite{Isberg:1993av,GCA,Bagchi:2009pe,Correlation,Galilean1,Galilean, Log,Alishahiha:2009np,Nishida:2007pj,Bagchi:2010vw,Hosseini:2015uba, Lukierski:2005xy, PhysRevD.5.377, 0264-9381-10-11-006, Martelli:2009uc,Duval:2009vt} through the \.In\"on\"u-Wigner contraction of the relativistic conformal algebra for $CFT_{1+1}$s.  
This contraction involves distinct scalings of the space and the time coordinates and hence breaks the Lorentz symmetry
 of the $CFT_{1+1}$ to a Galilean symmetry. The procedure modifies the original generators of the relativistic conformal algebra but leaves their numbers unchanged. The representation of the resulting Galilean conformal algebra (GCA) may then be utilized to determine the correlation functions of the primary fields in the $GCFT_{1+1}$. 

Interestingly the authors in \cite{Galilean1,Galilean} utilized a replica technique based on \cite {Calabrese:2004eu,Entanglement} to obtain the entanglement entropy of bipartite zero and finite temperature states in a $GCFT_{1+1}$.  As mentioned earlier entanglement negativity serves as a suitable measure for characterizing mixed state entanglement. This naturally
leads to the extremely interesting issue of determining the entanglement negativity for such bipartite states in a $GCFT_{1+1}$. In this article we address this significant issue and establish a construction involving a replica technique
similar to that described in \cite{Calabrese:2004eu,Entanglement,Galilean,Galilean1}, to obtain
the entanglement negativity for certain zero and finite temperature bipartite states in a $GCFT_{1+1}$. Interestingly 
our analysis reproduces the universal features of entanglement negativity for corresponding bipartite states of relativistic $CFT_{1+1}$s \cite{PhysRevLett.109.130502,Negativity,Calabrese:2014yza} in a $GCFT_{1+1}$ which strongly validates our construction.

This article is organized as follows. In section \ref{sec2}, we describe the entanglement negativity measure in quantum information theory and briefly review the replica technique for computing this quantity for various bipartite states in a relativistic $CFT_{1+1}$. Subsequently in section \ref{sec3}, the replica technique for computing the entanglement entropy of bipartite states in a $GCFT_{1+1}$ is reviewed. Following this in section \ref{sec4}, we compute the entanglement negativity for various bipartite pure and mixed states in a $GCFT_{1+1}$ employing the above replica technique. Finally in section \ref{sec5} we present a summary of our work and conclusions.

\section{Entanglement negativity}\label{sec2}

We begin by briefly reviewing the definition of entanglement negativity in quantum information theory \cite{Vidal} (see \cite{Rangamani:2014ywa} for a review). In this context we consider a tripartite system in a pure state constituted by the subsystems $A_1$, $A_2$ and $B$, where $A=A_1\cup A_2$ and $B=A^c$ represent the rest of the system. We assume that the Hilbert space $\mathcal{H}$ for the bipartite system $A$ factorizes as $\mathcal{H}=\mathcal{H}_1 \otimes \mathcal{H}_2$, where 
$\mathcal{H}_1, \mathcal{H}_2$ are the Hilbert spaces for the subsystems $A_1$ and $A_2$ respectively.
The partial transpose of the reduced density matrix for the bipartite system $A$ with respect to the second subsystem is defined as
\begin{equation}\label{26}
\left< e^{(1)}_ie^{(2)}_j|\rho_A^{T_2}|e^{(1)}_ke^{(2)}_l\right> = 
\left< e^{(1)}_ie^{(2)}_l|\rho_A|e^{(1)}_ke^{(2)}_j\right>,
\end{equation}
where $|e^{(1)}_i\rangle$ and $|e^{(2)}_j\rangle$ are the bases for the Hilbert spaces
$\mathcal{H}_1$ and  $\mathcal{H}_2$. The entanglement negativity for the bipartite mixed state configuration $A\equiv A_1 \cup A_2$ may then be defined as 
\begin{equation}\label{28}
\mathcal{E} = \ln \mathrm{Tr}||\rho_A^{T_2}||,
\end{equation}
where the trace norm $\mathrm{Tr}||\rho_A^{T_2}||$ is given by the sum of absolute eigenvalues of 
$\rho_A^{T_2}$.

\subsection{Entanglement negativity in a conformal field theory}

\begin{figure}[H]
\centering
\includegraphics[width=7cm]{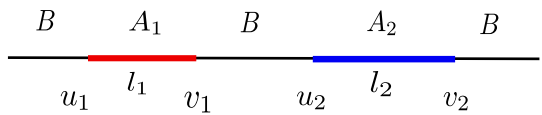}
\caption{Two disjoint intervals $A_1$ and $A_2$.}\label {tripart}
\end{figure}

As mentioned in the \nameref{sec1}, the entanglement negativity for bipartite states in a $CFT_{1+1}$ may be computed through a replica technique \cite{PhysRevLett.109.130502,Negativity}. For this purpose the tripartite system in a pure state is described by the spatial intervals $A_1\equiv [u_1,v_1]$, $A_2 \equiv [u_2,v_2]$ and $B$ as depicted in figure \ref {tripart}.

The entanglement negativity $\mathcal{E}$ for the bipartite mixed state configuration described by $A \equiv A_1\cup A_2$ may then be obtained by utilizing a replica technique involving the analytic continuation of the quantity $\mathrm{Tr}(\rho_A^{T_2})^{n_e}$ through even sequences of $n=n_e$  to $n_e\to1$ as follows

\begin{equation}\label{2.3}
\mathcal{E} = \lim_{n_e \rightarrow 1 } \ln \mathrm{Tr}(\rho_A^{T_2})^{n_e}.
\end{equation}

As described in \cite {PhysRevLett.109.130502,Negativity} the quantity on the right hand side of the above equation may be expressed as a four point twist correlator on the complex plane as
\begin{equation}\label{2.4}
\mathrm{Tr}(\rho_A^{T_2})^{n_e} = 
\langle\mathcal{T}_{n_e}(u_1)\overline{\mathcal{T}}_{n_e}(v_1)\overline{\mathcal{T}}_{n_e}(u_2)\mathcal{T}_{n_e}(v_2)\rangle_{\mathbb{C}}.
\end{equation}

\subsection{Negativity for a single interval}
The entanglement negativity for the pure state configuration of a single interval $A_1$ of length 
$l=u_2-v_2$ in an infinite system described by a $CFT_{1+1}$ may now be obtained by considering the limit $u_2\to v_1$, $v_2\to u_1$ in which the interval $A= A_1\cup A_2$ describes the full system with 
$B\to \emptyset$ where $\emptyset$ is the null set. In this limit the four point twist correlator in eq. (\ref {2.4}) reduces to the following two point twist correlator in the replica approach
\begin{equation}\label{2.5b}
    \textrm{Tr}(\rho_A^{T_2})^{n_e}=\left<\mathcal{T}_{n_e}^2(u_2)\overline{\mathcal{T}}_{n_e}^2(v_2)\right>.
\end{equation}
As explained in \cite {PhysRevLett.109.130502,Negativity} the $n_e$-sheeted Riemann surface decouples into two independent $(n_e/2)$-sheeted Riemann surfaces which leads to the following expression,
\begin{equation}\label{eq:(3.4)}
\textrm{Tr}(\rho_A^{T_2})^{n_e}=(\left<\mathcal{T}_{n_e/2}(u_2)\overline{\mathcal{T}}_{n_e/2}(v_2)\right>)^2=(\textrm{Tr}(\rho_{A_2}^{n_e/2}))^2.
\end{equation}
From eq. (\ref{eq:(3.4)}), the scaling dimensions of the twist fields $\mathcal{T}_{n_e}^2$ and $\bar{\mathcal{T}}_{n_e}^2$ are given as
\begin{equation}\label{2.7}
    \Delta_{n_e}^{(2)}=\frac{c}{6}\left (\frac{n_e}{2}-\frac{2}{n_e}\right ).
\end{equation}
Also the scaling dimension of twist fields $\mathcal{T}_{n_e}$ and $\bar{\mathcal{T}}_{n_e}$ is given by \cite{Calabrese:2004eu}
\begin{equation}
\Delta_{n_e}=\frac{c}{12}\left(n_e-\frac{1}{n_e}\right).
\end{equation}

The entanglement negativity for the pure state configuration of a single interval may then be obtained using eqs. (\ref{2.3}) and (\ref{eq:(3.4)}) in the replica limit $n_e\to 1$ as \cite{Negativity}
\begin{equation}
    \mathcal{E}=\frac{c}{2}\textrm{ln}\frac{l}{a}+2\ln c_{1/2} ,
\end{equation}
where $c_{1/2}$ is a non universal constant appearing in the two point twist correlator and $a$ is an UV cutof{}f. Note that the entanglement negativity for the pure state is described by the R\'{e}nyi entropy of order half as expected from quantum information theory. 
We observe that for a pure vacuum state, the relationship between the entanglement negativity and entanglement entropy\cite{Entanglement} of a single interval in the $CFT_{1+1}$ is given as
\begin{equation}\label{relation}
\mathcal{E}=\frac{3}{2}S_A + \mathrm{const}.
\end{equation}

As described in \cite{Negativity}, the entanglement negativity for the single interval in a finite system of length $L$ with periodic boundary condition may also be obtained from the corresponding two point twist correlator in eq. (\ref{2.5b}) on a cylinder of circumference $L$. In this case, the conformal map from the complex plane to the cylinder is given by $z\to w=\frac{i L}{2\pi}\ln z$. The entanglement negativity may then be obtained from eq. (\ref{2.3}) as 
\begin{equation}
\mathcal{E}=\frac{c}{2}\ln \left( \frac{L}{\pi a} \sin \frac{\pi l}{L}\right)+2\ln c_{1/2},
\end{equation}
where once again $c_{1/2}$ is a non universal constant and $a$ is a UV cutof{}f.

\subsection{Negativity for adjacent intervals}

Having described the entanglement negativity for the pure state configuration of a single interval we now focus our attention on a bipartite zero temperature mixed state configuration of adjacent intervals with lengths given as $v_1-u_1=l_1$, $v_2-u_2=l_2$. This is described by the limit $v_1\to u_2$ in which the four point twist correlator in 
eq. (\ref{2.4}) reduces to a three point twist correlator \cite{DiFrancesco:1997nk} on the complex plane  as follows
\begin{equation} \label{adjacent}
\textrm{Tr}(\rho_A^{T_2})^{n_e}=\left<\mathcal{T}_{n_e}(-l_1)\bar{\mathcal{T}}_{n_e}^2(0)\mathcal{T}_{n_e}(l_2)\right>=c_{n_e}^2\frac{C_{\mathcal{T}_{n_e}\bar{\mathcal{T}}_{n_e}^2\mathcal{T}_{n_e}}}{(l_1l_2)^{\Delta_{n_e}^{(2)}}(l_1+l_2)^{2\Delta_{n_e}-\Delta_{n_e}^{(2)}}},
\end{equation}
where $c_{n_e}$ and $C_{\mathcal{T}_{n_e}\bar{\mathcal{T}}_{n_e}^2\mathcal{T}_{n_e}}$ are constants appearing in the two point and three point twist correlators respectively. Insertion of the conformal weights for the twist operators in eq. (\ref{adjacent}) now leads to the following expression for the entanglement negativity of the mixed state under consideration \cite{Negativity}
\begin{equation}
    \mathcal{E}=\frac{c}{4}\textrm{ln}\frac{l_1l_2}{(l_1+l_2)a}+\textrm{const}.,
\end{equation}
where $a$ is a UV cutof{}f for the $CFT_{1+1}$.

The entanglement negativity for the mixed state of adjacent intervals in a finite sized system of length $L$ with a periodic boundary condition may now be obtained from the corresponding three point twist correlator in eq. (\ref{adjacent}) on a cylinder of circumference $L$. The conformal map from the complex plane to a cylinder is given by $z\to w=\frac{i L}{2\pi}\ln z$. The entanglement negativity in this case is then given as \cite{Negativity}
\begin{equation}
\mathcal{E}=\frac{c}{4}\ln \left[\left(\frac{L}{\pi a}\right)\frac{\sin(\frac{\pi l_1}{L})\sin (\frac{\pi l_2}{L})}{\sin \frac{\pi(l_1+l_2)}{L}}\right]+ \mathrm{const}.,
\end{equation}
where $a$ is a UV cutof{}f.

The corresponding entanglement negativity for the mixed state of adjacent intervals at a finite temperature $T=1/\beta$ involves the three point twist correlator on an infinite cylinder of circumference $\beta$ where the Euclidean time direction is now compactified. This may be determined through the conformal map from the complex plane to the cylinder given as $z\to w=\frac{\beta}{2\pi}\ln z$. This leads to the following expression for the entanglement negativity of the finite temperature mixed state as follows \cite{Negativity}
\begin{equation}
\mathcal{E}=\frac{c}{4}\ln \left[\left(\frac{\beta}{\pi a}\right)\frac{\sinh(\frac{\pi l_1}{\beta})\sinh (\frac{\pi l_2}{\beta})}{\sinh \frac{\pi(l_1+l_2)}{\beta}}\right]+ \mathrm{const}.,
\end{equation}
where $a$ is a UV cutof{}f.

\subsection{Negativity for a single interval at a finite temperature}\label{Sec2.4}
In \cite{Calabrese:2014yza} it was shown that the entanglement negativity for the case of a mixed state of a single interval at a finite temperature was much more subtle. As described there a naive computation of the entanglement negativity through the evaluation of the corresponding two point twist correlator on an infinite cylinder leads to an incorrect result due to subtle geometrical reasons. For this purpose it was necessary to consider the configuration of a single interval complemented by two other large but finite intervals adjacent to it which was described by a four point twist correlator. The entanglement negativity for the single interval could then be obtained through a bipartite limit in which the two other intervals are taken to be infinite, following the replica limit to arrive at
\begin{equation}\label{2.14}
    \mathcal{E}=\lim_{L\to \infty}\lim_{n_e\to1}\ln \left[\left<\mathcal{T}_{n_e}(-L)\overline{\mathcal{T}}^2_{n_e}(-l)\mathcal{T}^2_{n_e}(0)\overline{\mathcal{T}}_{n_e}(L)\right>_\beta\right].
\end{equation}
In the above equation $\mathcal{T}_{n_e}(\pm L)$ are the twist fields at the branch points located at the extremities of the auxiliary intervals adjacent on either side of the single interval and 
the subscript $\beta$ denotes that the four point twist correlator is evaluated on a cylinder of circumference $\beta=\frac{1}{T}$. It is possible to determine the four point twist correlator described in eq. (\ref {2.14})
on the complex plane up to a non universal function $\mathcal{F}_{n_e}(x)$ of the cross ratio in both the limits $x\to 1$ and $x\to 0$ as
\begin{equation}\label{2.16}
\left<\mathcal{T}_{n_e}(z_1)\overline{\mathcal{T}}^2_{n_e}(z_2)\mathcal{T}^2_{n_e}(z_3)\overline{\mathcal{T}}_{n_e}(z_4)\right>_\mathbb{C}=\frac{c_{n_e}c^2_{n_e/2}}{z_{14}^{2\Delta_{n_e}}z_{23}^{2\Delta_{n_e}^{(2)}}}\frac{\mathcal{F}_{n_e}(x)}{x^{\Delta_{n_e}^{(2)}}}, \,\,\,\,\, x=\frac{z_{12}z_{34}}{z_{13}z_{24}}.
\end{equation}
Here $\Delta_{n_e}$, $\Delta_{n_e}^{(2)}$ are the scaling dimensions of $\mathcal{T}_{n_e}$ and $\mathcal{T}^2_{n_e}$ respectively. As described in \cite{Calabrese:2014yza} the non universal arbitrary function $\mathcal{F}_{n_e}(x)$ at the limits $x\to 1$ and $x\to 0$ is given as
\begin{equation}
\mathcal{F}_{n_e}(1)=1, \hspace{5mm} \mathcal{F}_{n_e}(0)=C_{n_e},
\end{equation}
where $C_{n_e}$ is a constant depending upon the full operator content of the theory. It is now possible to evaluate the four point twist correlator on a cylinder of circumference $\beta$ through the standard conformal map from the complex plane to the cylinder via $z\to w=\beta/2\pi \,\ln z$.
This leads us to the entanglement negativity of a single interval at a finite temperature as follows \cite{Calabrese:2014yza}
\begin{equation} \label{2.17}
\begin{aligned}
\mathcal{E}&=\frac{c}{2}\ln\Big[\frac{\beta}{\pi a}\sinh\left(\frac{\pi l}{\beta}\right)\Big]-\frac{\pi c l}{2\beta}+f(e^{-2\pi l/\beta})+\mathrm{const.},
\end{aligned}
\end{equation}
where $ \displaystyle f(x) = \lim_{n_e \to 1} \ln \left [ \mathcal{F}_{n_e}(x) \right ] $.
Note that the first two terms in the above expression are universal and the other terms involving the function $f(e^{-2\pi l/\beta})$ and the constant are non universal. Interestingly eq. (\ref {2.17}) may be expressed in the following intriguing form 
\begin{equation}
\begin{aligned}\label{2.19}
\mathcal{E}&=\frac{3}{2} \big[ S_A-S_A^{th} \big] +f(e^{-2\pi l/\beta})+\mathrm{const.},
\end{aligned}
\end{equation}
where $S_A$ and $S_A^{th}$ denote the entanglement entropy and the thermal entropy of the finite temperature mixed state in question respectively. The above expression demonstrates that the universal part of the entanglement negativity is described by the elimination of the thermal entropy from the entanglement entropy for the finite temperature mixed state, justifying its characterization as the upper bound on the {\it distillable entanglement} in quantum information theory.

\section{Entanglement entropy in a Galilean conformal field theory}\label{sec3}

In this section we briefly recapitulate the salient features of a class of non relativistic $CFT_{1+1}$ with Galilean invariance ($GCFT_{1+1}$) \cite{Isberg:1993av,GCA,Bagchi:2009pe,Correlation,Galilean, Galilean1} and describe the characterization of the entanglement entropy in these theories.
The $GCFT_{1+1}$ above involves the Galilean conformal algebra (GCA) which may be obtained from the usual relativistic conformal algebra through an \.In\"on\"u-Wigner contraction that necessitates a rescaling of the space and time coordinates as follows
\begin{equation}
    t\to t,\quad   x_i\to \epsilon x_i,
\end{equation}
with $\epsilon\to 0$. This is equivalent to considering the velocity $v_i\sim\epsilon$ going to zero. The generators of the (1+1)-dimensional GCA in the plane representation \cite{Galilean1} are given as\footnote{Note that we are working in the plane representation which differs from the normal representation used in \cite{Correlation,Bagchi:2009pe} by a negative sign in the GCA.}
\begin{equation}
L_n=t^{n+1}\partial_t+(n+1)t^nx\partial_x, \quad M_n=t^{n+1}\partial_x,
\end{equation}
which leads to the Lie algebra with a central extension as
\begin{equation}
\begin{aligned}
    \left[L_n,L_m\right]&= (m-n)L_{n+m}+\frac{c_L}{12}(n^3-n)\delta_{n+m,0},
    \\ [L_n,M_m]&=(m-n)M_{n+m}+\frac{c_M}{12}(n^3-n)\delta_{n+m,0},\\
    [M_n,M_m]&=0.
\end{aligned}
\end{equation}
Here the quantities $c_L$ and $c_M$ are the central charges for the GCA.\footnote{Note that the two central charges $c_L$ and $c_M$ are the analogues of the central charges $(c, {\bar c})$ for the two copies of the conformal algebra in a relativistic $CFT_{1+1}$.}

A state in the $GCFT_{1+1}$ for the highest weight representation of the GCA \cite{Correlation,Bagchi:2009pe}, is labeled by the conformal weights $h_L$ and $h_M$ given as 
\begin{equation}
L_0\ket{h_L,h_M} = h_L\ket{h_L,h_M}, \qquad M_0\ket{h_L,h_M}= h_M\ket{h_L,h_M}.
\end{equation}
The two point correlator for primary fields $V(x,t)$ in a $GCFT_{1+1}$ may be determined from the Galilean conformal symmetry characterized by the GCA as follows \cite{Correlation}
\begin{equation}\label{3.6}
   \big <V_1(x_1,t_1)V_2(x_2,t_2)\big >=C^{(2)}\delta_{h_L^{1}h_L^{2}}
   \delta_{h_M^{1}h_M^{2}}t_{12}^{-2h_L^{1}}\exp\left(-2h_M^{1}\frac{x_{12}}{t_{12}}\right).
\end{equation}
Here $(h_L^{1},h_M^{1})$ and $(h_L^{2},h_M^{2})$ are the weights of the primary fields $V_1$ and $V_2$ respectively, $C^{(2)}$ is a normalization constant and
$x_{12}=x_1-x_2,\, t_{12}=t_1-t_2$ . In an exactly similar fashion it is also possible to determine the three point function of primary fields in a $GCFT_{1+1}$ to be as follows \cite{Correlation}
\begin{equation}\label{eq:(4.5)}
\begin{aligned}
\left<V_1(x_1,t_1)V_2(x_2,t_2)V_3(x_3,t_3)\right>&=C^{(3)}t_{12}^{-(h_L^1+h_L^2-h_L^3)}\,
    t_{23}^{-(h_L^2+h_L^3-h_L^1)}\,
   t_{13}^{-(h_L^1+h_L^3-h_L^2)}\times\\
   &\exp\Big[-(h_M^1+h_M^2-h_M^3)\frac{x_{12}}{t_{12}} 
   -(h_M^2+h_M^3-h_M^1)\frac{x_{23}}{t_{23}}\\
   &-(h_M^1+h_M^3-h_M^2)\frac{x_{13}}{t_{13}}\Big].
\end{aligned}
\end{equation}
Here the $V_i$'s  are primary fields with weights $\{(h_L^i,h_M^i)\}$ and $x_{ij}=x_i-x_j$, $t_{ij}=t_i-t_j$ with $(i=1,2,3)$ respectively and $C^{(3)}$ is a constant.
Similarly, the four point function of primary fields in the $GCFT_{1+1}$ may be expressed as 
\begin{equation}\label{3.7c}
\begin{aligned}
\left<\prod_{i=1}^4 V_i(x_i,t_i)\right>&=\frac{t_{13}^{h_L^1+h_L^3}t_{24}^{h_L^2+h_L^4}}{t_{12}^{h_L^1+h_L^2}t_{23}^{h_L^2+h_L^3}t_{34}^{h_L^3+h_L^4}t_{14}^{h_L^1+h_L^4}}
 \exp\Big [\frac{x_{13}}{t_{13}}(h_M^1+h_M^3)
 +\frac{x_{24}}{t_{24}}(h_M^2+h_M^4)\\
 &-\frac{x_{12}}{t_{12}}(h_M^1+h_M^2)-\frac{x_{23}}{t_{23}}(h_M^2+h_M^3)-\frac{x_{34}}{t_{34}}(h_M^3+h_M^4)-\frac{x_{14}}{t_{14}}(h_M^1+h_M^4)\Big ]\,\,  \mathcal{G}(t,\frac{x}{t}),
 \end{aligned}
\end{equation}
where $\{(h_L^i,h_M^i)\}$ are the weights of the primary fields $V_i(x_i,t_i)$ respectively with $(i=1,2,3,4)$ and
\begin{equation}\label{4.15b}
t=\frac{t_{12}t_{34}}{t_{13}t_{24}}, \hspace{5mm}\frac{x}{t}=\frac{x_{12}}{t_{12}}+\frac{x_{34}}{t_{34}}-\frac{x_{13}}{t_{13}}-\frac{x_{24}}{t_{24}},
\end{equation}
which are the non relativistic analogues of the cross ratio $x$ described in subsection \ref{Sec2.4}. Note that the function $ \mathcal{G}(t,\frac{x}{t})$ is a non universal function of the cross ratios which depend on the specific operator content of the $GCFT_{1+1}$.

\subsection{Entanglement entropy of a single interval at zero temperature}
\begin{figure}[H]
\centering
\includegraphics[width=6cm]{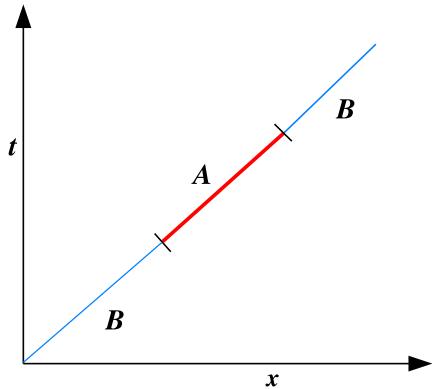}
\caption{Galilean boosted interval A and B describes the rest of the system \cite{Galilean}.}\label{fig2}
\end{figure}

Having described the two and the three point correlation functions of the primary fields in a $GCFT_{1+1}$, we now briefly review the replica technique for the computation of the entanglement entropy of a bipartite state described by a single interval in a $GCFT_{1+1}$ \cite{Galilean,Galilean1}. The computation of the entanglement entropy in a $GCFT_{1+1}$ is similar to that in a relativistic $CFT_{1+1}$ with certain modifications. Unlike a relativistic $CFT_{1+1}$ where it was possible to utilize a fixed time slice due to the Lorentz invariance, in a $GCFT_{1+1}$ it is required to consider an interval with a Galilean boost. Such a boosted interval $A$ may be described by the end points $(x_1,t_1)$ and $(x_2,t_2)$, as shown in the figure \ref {fig2} where the complement $B=A^c$ denotes the rest of the system.
In this case the partition function $Z_n(A)$ is defined on an $n$-sheeted surface 
$\Sigma_n$, consisting of $n$ copies of the GCFT  plane $(x,t)$ glued together. The corresponding transformation from the GCFT  plane $(x,t)$ to the $(x',t')$ coordinates on the
$n$-sheeted surface $\Sigma_n$, is given as \cite{Galilean}
\begin{equation}
\begin{aligned}
t = \left(\frac{t'-t_1}{t'-t_2}\right)^{1/n},\qquad \qquad
x = \frac{1}{n}\left(\frac{t'-t_1}{t'-t_2}\right)^{1/n}\left(\frac{x'-x_1}{t'-t_1}-\frac{x'-x_2}{t'-t_2}\right).
\end{aligned}
\end{equation}
The entanglement entropy for the bipartite pure state $A$ of a single interval in the $GCFT_{1+1}$ may then be characterized through the product of $n$ two point twist field correlators on the GCFT plane as follows \cite {Galilean}
\begin{equation}
\begin{aligned}
    \textrm{Tr}\rho_A^n&=k_n\left(\left<\Phi_n(x_1,t_1)\Phi_{-n}(x_2,t_2)\right>\right)^n\\
    &=k_nt_{12}^{-\frac{c_L}{12}(n-\frac{1}{n})}\exp\left[-\frac{c_M}{12}\left(n-\frac{1}{n}\right)\frac{x_{12}}{t_{12}}\right].
    \end{aligned}
\end{equation}
In the above expression $\Phi_n(x_1,t_1)$ and $\Phi_{-n}(x_2,t_2)$ are the twist and the anti twist fields localized at the end points of the interval $A$ with weights given as $h_L=\frac{c_L}{24}(1-\frac{1}{n^2})$ and
$h_M=\frac{c_M}{24}(1-\frac{1}{n^2})$ which may be determined from the $GFCT_{1+1}$ conformal Ward identities and $k_n$ are some normalization constants. Utilizing the above expression the entanglement entropy for the bipartite pure state described by a single interval in the $GCFT_{1+1}$ may then be obtained as \cite{Galilean}
\begin{equation}\label{eq:4.8}
    S_A=-\lim_{n\to1}\frac{\partial}{\partial n}\textrm{Tr}\rho_A^n=\frac{c_L}{6}\textrm{ln}\left(\frac{t_{12}}{a}\right)+\frac{c_M}{6}\left(\frac{x_{12}}{t_{12}}\right),
\end{equation}
where $a$ is a UV cutof{}f.

\subsection{Entanglement entropy at a finite temperature}

For the corresponding finite temperature bipartite mixed state of the single interval denoted by $A$ the entanglement entropy is characterized by the two point twist correlator for the $GCFT_{1+1}$ on a cylinder with circumference $\beta=\frac{1}{T}$ where $T$ is the temperature in a $GCFT_{1+1}$  \cite{Calabrese:2004eu}.\footnote{Note that unlike in a relativistic $CFT_{1+1}$ the holomorphic and the antiholomorphic sectors do not decouple in a $GCFT_{1+1}$. Hence a clear identification of the coordinates as time like or space like on the cylinder is not directly evident. The consistency of the construction is validated by explicit limits.} The transformation law for a  primary field $\phi$ in relativistic $CFT_{1+1}$ from the CFT plane to a cylinder $(z\to w=\frac{\beta}{2\pi}\ln z)$ is given as
\begin{equation}\label{cftprimary}
\tilde{\phi}(\omega,\bar{\omega})=\left(\frac{d\omega}{dz}\right)^{-h}\left(\frac{d\bar{\omega}}{d\bar{z}}\right)^{-\bar{h}}\phi(z,\bar{z}).
\end{equation}
An \.In\"on\"u-Wigner contraction ($t\to t, x\to\epsilon x$) of the above coordinate transformation $z\to w=\frac{\beta}{2\pi}\ln z$ where the coordinates in the Lorentzian signature are described by $z=t+x$ and $\bar{z}=t-x$, leads to the conformal map from the coordinates $(x, t)$ on the GCFT plane to the coordinates $(\xi, \rho)$ on the cylinder as follows \cite{Log}
\begin{equation}\label{eq:(3.10)}
t=e^{\frac{2\pi\xi}{\beta}}, \,\,\,\, x=\frac{2\pi\rho}{\beta}e^{\frac{2\pi\xi}{\beta}}.
\end{equation}
A similar contraction of eq. \eqref {cftprimary} provides the appropriate transformation of the GCFT primary fields from the $GCFT$ plane to the cylinder as follows \cite{Log}
\begin{equation}\label{eq:(3.11)}
\tilde{\Phi}(\xi,\rho)=\left(\frac{\beta}{2\pi}\right)^{-h_L} e^{\frac{2\pi}{\beta}(\xi h_L+\rho h_M)}\Phi(x,t).
\end{equation}

From the above transformations for the primary fields in the $GCFT_{1+1}$ given in eq. (\ref{eq:(3.11)}), the two point twist correlator on the cylinder may be expressed as follows 
\cite { Galilean1}
\begin{equation}\label{3.14}
\left<\Phi_n(\xi_1,\rho_1)\Phi_{-n}(\xi_2,\rho_2)\right >=\left[\frac{\beta}{\pi}\mathrm{sinh}\left(\frac{\pi\xi_{12}}{\beta}\right)\right]^{-2h_L} \exp\left[-2h_M\frac{\pi\rho_{12}}{\beta}\mathrm{coth}\left(\frac{\pi\xi_{12}}{\beta}\right)\right].
\end{equation}
The entanglement entropy for the finite temperature mixed state of a single interval described by the end points $(\xi_1,\rho_1)$ and $(\xi_2,\rho_2)$ on the cylinder in the $GCFT_{1+1}$ is then obtained from eq. (\ref{3.14}) as follows \cite{Galilean1,Galilean}
\begin{equation}\label{3.14a}
S_A=\frac{c_L}{6}\mathrm{ln}\left[\frac{\beta}{\pi a}\mathrm{sinh}\left(\frac{\pi\xi_{12}}{\beta}\right)\right]+\frac{\pi c_M}{6\beta}\rho_{12}\coth\left(\frac{\pi\xi_{12}}{\beta}\right).
\end{equation}
The limit $\beta \to \infty$ for the above expression leads to the earlier zero temperature result for the entanglement entropy of the pure state described in eq. \eqref {eq:4.8} which validates the consistency of the result.

\subsection{Entanglement entropy for a finite size system}
The entanglement entropy for the zero temperature pure state of a single interval in a finite system of length $L$ with a periodic boundary condition may be computed from the corresponding two point twist correlator for the $GCFT_{1+1}$ on a cylinder of circumference $L$. The relevant conformal map from the GCFT plane described by the coordinates $(x,t)$ to the cylinder in this case is given as follows
\begin{equation}\label{eq:(3.14)}
t=e^{\frac{2\pi i\xi}{L}}, \,\,\,\, x=\frac{2\pi i\rho}{L}e^{\frac{2\pi i\xi}{L}}.
\end{equation}
where $(\xi, \rho)$ are the coordinates on the cylinder. The GCFT primaries under this transformation transforms as
\begin{equation}\label{3.18f}
\tilde{\Phi}(\xi,\rho)=\left(\frac{L}{2\pi i}\right)^{-h_L} e^{\frac{2\pi i}{L}(\xi h_L+\rho h_M)}\Phi(x,t).
\end{equation}
Using the above transformation of fields, the two point twist correlator on the cylinder is described as
\begin{equation}\label{3.17}
\left<\Phi_n(\xi_1,\rho_1)\Phi_{-n}(\xi_2,\rho_2)\right >
=\left[\frac{L}{\pi}\mathrm{sin}\left(\frac{\pi\xi_{12}}{L}\right)\right]^{-2h_L} \exp\left[-2h_M\frac{\pi\rho_{12}}{L}\mathrm{cot}\left(\frac{\pi\xi_{12}}{L}\right)\right].
\end{equation}
As earlier the entanglement entropy for the zero temperature pure state of a single interval with end points at $(\xi_1,\rho_1)$ and $(\xi_2,\rho_2)$
in a finite sized system described by the $GCFT_{1+1}$ on a spatial cylinder is then given from eq. (\ref{3.17}) as follows \cite{Galilean,Galilean1}
\begin{equation}\label{3.17A}
S_A=\frac{c_L}{6}\mathrm{ln}\left[\frac{L}{\pi a}\mathrm{sin}\left(\frac{\pi\xi_{12}}{L}\right)\right]+\frac{\pi c_M}{6L}\rho_{12}\cot\left(\frac{\pi\xi_{12}}{L}\right).
\end{equation}
where $a$ is a UV cutof{}f and $\rho_{12}=\mid \rho_1-\rho_2 \mid $. The limit $L\to\infty$ for the above expression once again leads to eq. (\ref{eq:4.8}) which refers to an infinite system validating the result.

\section{Entanglement negativity in a $GCFT_{1+1}$ }\label{sec4}

\begin{figure}[H]
\centering
\includegraphics[width=7cm]{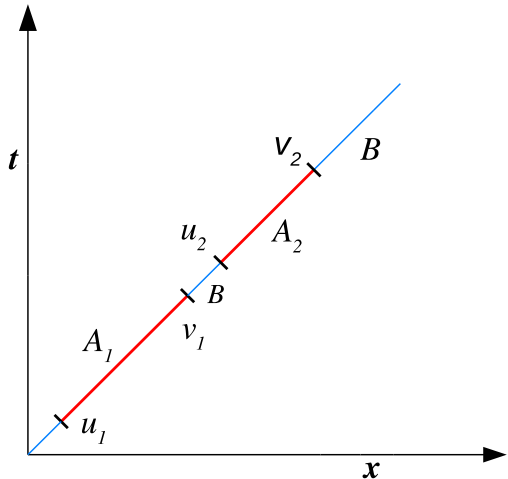}
\caption{Intervals $A_1$ and $A_2$ in the GCFT plane.}\label{fig3}
\end{figure}

Having reviewed the computation of the entanglement entropy for various zero and finite temperature bipartite pure and mixed state configurations in a $GCFT_{1+1}$, in this section we now turn our attention to the definition of the entanglement negativity for such bipartite states. To this end we consider the configuration described by the intervals 
$A=A_1\cup A_2$ where $A_1\equiv [u_1,v_1], A_2\equiv [u_2,v_2]$ and $B=A^c$ describes the rest of the system as depicted in figure \ref{fig3} . Here $u_1\equiv (x_1,t_1)$, $v_1\equiv (x_2,t_2), u_2\equiv (x_3,t_3)$, $v_2\equiv (x_4,t_4)$ are the end points of the intervals $A_1, A_2$ respectively.

Following the procedure described in \cite{Negativity} it is possible to define the
partial transpose of the reduced density matrix $\rho_A$ with respect to the second interval $A_2$ which corresponds to the exchange of the row and the column indices 
for the interval $A_2$. The replica technique to compute the trace norm of the partially transposed reduced density matrix follows from that described in section \ref{sec2}. Accordingly in the present case the twist fields at the end points of the second interval are reversed in the four point correlator to arrive at
\begin{equation}\label{eq:(5.1)}
		\textrm{Tr}(\rho_A^{T_2})^{n_e}=\left<\Phi_{n_e}(u_1)\Phi_{-n_e}(v_1)\Phi_{-n_e}(u_2)\Phi_{n_e}(v_2)\right> ,
\end{equation}
where $\Phi_{\pm n_e}$ are the twist fields in the $GCFT_{1+1}$ and the replica index $n_e$ is even. As described in  eq. (\ref{2.3}) for the relativistic $CFT_{1+1}$, the entanglement negativity for the bipartite state depicted by the configuration of disjoint intervals in the $GCFT_{1+1}$ may also be defined through the replica limit $n_e\to 1$ as follows
\begin{equation}\label{eq: (5.1A}
\mathcal{E} = \lim_{n_e \rightarrow 1 } \ln \mathrm{Tr}(\rho_A^{T_2})^{n_e}.
\end{equation}

\subsection{Negativity for a single interval}
Utilizing the expression in eq. (\ref{eq: (5.1A}) given above we may now obtain the entanglement negativity for the pure vacuum state configuration of a single interval in the $GCFT_{1+1}$ through the bipartite limit of 
$B\to \emptyset$ described by $u_2\to v_1$ and $v_2\to u_1$ where the interval $A_2$ now describes the rest of the system. In this limit the four point twist correlator in eq. (\ref{eq:(5.1)}) reduces to a two point twist correlator as follows
\begin{equation}\label{eq:(5.2)}
    \textrm{Tr}(\rho_A^{T_2})^{n_e}=\left<\Phi_{n_e}^2(u_1)\Phi_{-n_e}^2(v_1)\right>.
\end{equation}

As described in \cite{Negativity} the $n_e$-sheeted surface 
$\Sigma_{n_e}$ decouples into two independent $(n_e/2)$-sheeted surfaces in this case and
hence in this limit the two point twist correlator in eq. (\ref{eq:(5.2)}) reduces to the following
\begin{equation}\label{eq:(5.3)}
\begin{aligned}
    \textrm{Tr}(\rho_A^{T_2})^{n_e}&=(\left<\Phi_{n_e/2}(u_1)\Phi_{-n_e/2}(v_1)\right>)^2=(\textrm{Tr}(\rho_{A_1}^{n_e/2}))^2.
\end{aligned}
\end{equation}
The corresponding weights of the twist fields $\Phi_{\pm n_e}$ may then be determined as described in section \ref{sec3} to be $\Delta_{n_e}=\frac{c_L}{24}\left(n_e-\frac{1}{n_e}\right)$ and $\chi_{n_e}=\frac{c_M}{24}\left(n_e-\frac{1}{n_e}\right)$.\footnote{These are related to the weights $h_L$ and $h_M$ as $\Delta_{n_e}=n_eh_L$ and $\chi_{n_e}=n_eh_M$ respectively \cite {Galilean}.}
Comparing (\ref{eq:(5.2)}) and (\ref{eq:(5.3)}), the weights of twist fields $\Phi_{\pm n_e}^2$ are determined as
\begin{equation}\label{eq:(5.4)}
\begin{aligned}
\Delta_{n_e}^{(2)}=\frac{c_L}{12}\left(\frac{n_e}{2}-\frac{2}{n_e}\right),
\\
\chi_{n_e}^{(2)}=\frac{c_M}{12}\left(\frac{n_e}{2}-\frac{2}{n_e}\right).
\end{aligned}
\end{equation}
Upon substitution of the weights for the twist fields from eq. (\ref{eq:(5.4)}) into eq. (\ref{eq:(5.2)}), we may express the left hand side of eq. (\ref{eq:(5.2)}) as follows
\begin{equation}
\textrm{Tr}(\rho_A^{T_2})^{n_e}
=k_{{\frac{n_e}{2}}}^2 t_{12}^{-\frac{c_L}{6}\left(\frac{n_e}{2}-\frac{2}{n_e}\right)}\exp\left[-\frac{c_M}{6}\left(\frac{n_e}{2}-\frac{2}{n_e}\right)\frac{x_{12}}{t_{12}}\right].
\end{equation}
The analytic continuation $n_e\to 1$ leads us to the trace norm of the reduced density matrix describing the pure state of a single interval in a $GCFT_{1+1}$ as
\begin{equation}
\begin{aligned}
    ||\rho_A^{T_2}||&=\lim_{n_e\to 1}\textrm{Tr}(\rho_A^{T_2})^{n_e}\\
    &=k_{1/2}^2\,t_{12}^{\frac{c_L}{4}}\exp\left(\frac{c_M}{4}\frac{x_{12}}{t_{12}}\right)
 \end{aligned}
\end{equation}
where $x_{12},t_{12}$ are as defined earlier. Finally, the entanglement negativity for the bipartite pure state configuration of a single interval $A$ in a $GCFT_{1+1}$ may be obtained from the above expression as follows
\begin{equation}\label{4.8n}
\begin{aligned}
\mathcal{E}&=\ln ||\rho_A^{T_2}||\\
&=\frac{c_L}{4}\ln\left(\frac{t_{12}}{a}\right)+\frac{c_M}{4}\left(\frac{x_{12}}{t_{12}}\right)+2\,\mathrm{ ln }\,k_{1/2}.
\end{aligned}
\end{equation}
Note that the first two terms on the right hand side of the above equation are universal whereas the last term is a non universal constant that depends on the specific $GCFT_{1+1}$ being considered.
Interestingly from eq. (\ref{4.8n}) we observe that the entanglement negativity for the bipartite pure state of a single interval in a $GCFT_{1+1}$ is given by the R\'{e}nyi entropy of order half similar to that of a relativistic $CFT_{1+1}$. From (\ref{eq:4.8}) for the entanglement entropy of the pure state of a single interval in a $GCFT_{1+1}$ the above equation may be expressed in the following form
\begin{equation}
\mathcal{E}=\frac{3}{2}S_A + \mathrm{const}.
\end{equation}
Note that the form of the universal part of the entanglement negativity is identical to the corresponding form in a relativistic $CFT_{1+1}$ as given in eq. (\ref{relation}). This is extremely significant as our construction reproduces the universal features of entanglement negativity in relativistic $CFT_{1+1}$s illustrating that these are determined purely by the conformal symmetry. Naturally this also serves as a strong consistency check for the validity of our construction.

\subsection{Entanglement negativity of a single interval for finite size systems}
It is now possible to extend the above results for the entanglement negativity 
for the bipartite pure state of a single interval in a $GCFT_{1+1}$ to a finite size system of length $L$ with a periodic boundary condition. For this purpose we employ the transformation of the primary fields given in eq. (\ref{3.18f})  to express the two point twist correlator on the cylinder as follows
\begin{equation}\label{4.10b}
\begin{aligned}
\left<\Phi^2_{n_e}(\xi_1,\rho_1)\Phi^2_{-n_e}(\xi_2,\rho_2)\right>
&=\left(\frac{L}{2\pi i}\right)^{-2\Delta_{n_e}^{(2)}} \exp\Bigg[\frac{2\pi i}{L}(\xi_1\Delta_{n_e}^{(2)}+\xi_2\Delta_{n_e}^{(2)}
+\rho_1\chi_{n_e}^{(2)}+\rho_2\chi_{n_e}^{(2)})\Bigg]\\
&\times t_{12}^{-2\Delta_{n_e}^{(2)}}\exp \left(-2\chi_{n_e}^{(2)}\frac{x_{12}}{t_{12}}\right),
\end{aligned}
\end{equation}
where $(\Delta_{n_e}^{(2)},\chi_{n_e}^{(2)})$  are the weights of $\Phi^2_{\pm n_e}$.
Utilizing the transformations described in eq. (\ref{eq:(3.14)}) to eq. (\ref{4.10b}), the two point twist correlator on the cylinder is given as
\begin{equation}
\left<\Phi^2_{n_e}(\xi_1,\rho_1)\Phi^2_{-n_e}(\xi_2,\rho_2)\right>
=\left[\frac{L}{\pi}\sin\left(\frac{\pi \xi_{12}}{L}\right)\right]^{-2\Delta_{n_e}^{(2)}}
\exp\left[-2\chi_{n_e}^{(2)}\frac{\pi \rho_{12}}{L}\cot\left(\frac{\pi \xi_{12}}{L}\right)\right].
\end{equation}
Substituting the weights of twist fields from eq. (\ref{eq:(5.4)}) in above expression and using eq. (\ref{eq: (5.1A}), we arrive at the entanglement negativity for the bipartite pure state of a single interval in a finite sized system described by a $GCFT_{1+1}$ on a cylinder as follows
\begin{equation}
\mathcal{E}=\frac{c_L}{4}\ln\left[\frac{L}{\pi a}\sin\left(\frac{\pi\xi_{12}}{L}\right)\right]+\frac{\pi c_M}{4L}\rho_{12}\cot\left(\frac{\pi\xi_{12}}{L}\right)+ \, \mathrm{const}.
\end{equation}
Once again we observe that the entanglement negativity for the pure state of a single interval in a finite sized system described by the $GCFT_{1+1}$ is given by the R\'{e}nyi entropy of order half and the universal part is proportional to the entanglement entropy given in (\ref{3.17A}) through the following expression
\begin{equation}
\mathcal{E}=\frac{3}{2}S_A+\mathrm{const}.
\end{equation}

\subsection{Negativity for a single interval at a finite temperature}
We now turn to the issue of obtaining the entanglement negativity for the finite temperature bipartite mixed state of a single interval following the analysis described for a relativistic $CFT_{1+1}$ in \cite{Calabrese:2014yza} and reviewed in section \ref{sec2}. As earlier the entanglement negativity for the finite temperature mixed state of a single interval in a $GCFT_{1+1}$ involves a four point twist correlator on an infinite cylinder arising from the configuration of a single interval sandwiched between two adjacent large but finite intervals. The entanglement negativity for the mixed state in question may then be obtained as earlier, through a bipartite limit subsequent to the replica limit leading to the following expression
\begin{equation}
\mathcal{E}= \lim_{L\to \infty} \lim_{n_{e}\to 1}
\ln\left[\left<\Phi_{n_e}(-L,-y)\,\Phi^2_{-n_e}(-\xi,-\rho)\,\Phi^2_{n_e} (0,0)\,\Phi_{-n_e}(L,y)\right>_\beta\right].
\end{equation}
In the above expression the fields $\Phi_{\mp n_e} (\pm L, \pm y)$ are the twist fields located at the extremities of the auxiliary intervals adjacent to the single interval and the fields $\Phi^2_{-n_e} (-\xi,-\rho)$, $\Phi^2_{n_e} (0,0)$ are the twist fields located at the ends of the single interval such that $L>\xi>0$ and $y>\rho>0$. The coordinates $(\xi , \rho)$ are defined on the infinite cylinder of circumference $\beta =1/T$ where $T$ is the temperature. 

Now employing eq. (\ref{3.7c}) we may describe the four point twist correlator on the GCFT plane as follows
\begin{equation}\label{4.17c}
\begin{aligned}
\left<\Phi_{n_e}(x_1,t_1)\,\Phi^2_{-n_e}(x_2,t_2)\,\Phi_{n_e}^2(x_3,t_3)\,\Phi_{-n_e}(x_4,t_4)\right>
&=\frac{1}{t_{14}^{2\Delta_{n_e}}t_{23}^{2\Delta_{n_e}^{(2)}}t^{\Delta_{n_e}+\Delta_{n_e}^{(2)}}} 
\exp \Big [ - 2 \chi_{n_e} \frac{x_{14}}{t_{14}} \\
&- 2 \chi_{n_e}^{(2)} \frac{x_{23}}{t_{23}}
- (\chi_{n_e}+\chi_{n_e}^{(2)} ) \frac{x}{t} \Big ] \,\mathcal{G}_{n_e}(t,\frac{x}{t}).
\end{aligned}
\end{equation}
In the above equation $\mathcal{G}_{n_e}(t,\frac{x}{t})$ is an arbitrary non universal function of the cross ratios.
This may be fixed in certain limits of the cross ratios $t$ and $\frac{x}{t}$ which are related to each other. Hence it suffices to consider the function $\mathcal{G}_{n_e}(t,\frac{x}{t})$ in the limits
$t\to 1$ and $t\to 0$ and express the four point twist correlator in eq. (\ref{4.17c}) in these limits
as follows
\begin{equation}\label{4.30f}
\begin{aligned}
\left<\Phi_{n_e}(x_1,t_1)\,\Phi^2_{-n_e}(x_2,t_2)\,\Phi_{n_e}^2(x_3,t_3)\,\Phi_{-n_e}(x_4,t_4)\right>
&=\frac{k_{n_e} \,k_{n_e/2}^2}{t_{14}^{2\Delta_{n_e}}\,t_{23}^{2\Delta_{n_e}^{(2)}}}\frac{\mathcal{F}_{n_e}(t,x/t)}{t^{\Delta_{n_e}^{(2)}}}\\
& \times \exp \Bigg [ - 2 \chi_{n_e} \frac{x_{14}}{t_{14}}
- 2 \chi_{n_e}^{(2)} \frac{x_{23}}{t_{23}} - \chi_{n_e}^{(2)} \frac{x}{t} \Bigg ].
\end{aligned}
\end{equation}
Here the non universal function $\mathcal{F}_{n_e}(t,x/t)$ of the cross ratios which is related to the function $\mathcal{G}_{n_e}(t,\frac{x}{t})$, in the limits $t\to 1$ and $t\to 0$ is given as
\begin{equation}
\mathcal{F}_{n_e}(1,0)=1, \hspace{5mm}  \mathcal{F}_{n_e}(0,\frac{x}{t})=C_{n_e},
\end{equation}
where $C_{n_e}$ is a constant that depends on the full operator content of the theory.
We now utilize  the conformal map from the $GCFT_{1+1}$ plane to the cylinder given by eq. (\ref{eq:(3.10)}) and the corresponding field transformations from eq. (\ref{eq:(3.11)}) to express the four point function on the cylinder in the following way
\begin{equation}\label{4.20c}
\begin{aligned}
\left<\Phi_{n_e}(-L,-y)\,\Phi^2_{-n_e}(-\xi,-\rho)\,\Phi^2_{n_e} (0,0)\,\Phi_{-n_e}(L,y)\right>_\beta=\frac{k_{n_e} \, k_{n_e/2}^{2}}{t^{\Delta_{n_e}^{(2)}}}\left [\frac{\beta}{\pi}\sinh\left( \frac{2\pi L}{\beta}\right )\right ]^{-2\Delta_{n_e}}\\
\left [\frac{\beta}{\pi}\sinh \left(\frac{\pi\xi}{\beta}\right)\right ]^{-2\Delta_{n_e}^{(2)}}\exp\left[-\frac{2\pi y}{\beta}\coth\left(\frac{2\pi L}{\beta}\right)2\chi_{n_e}-\frac{2\pi \rho}{\beta}\coth\left(\frac{\pi\xi}{\beta}\right)\chi_{n_e}^{(2)}-\frac{x}{t}\chi_{n_e}^{(2)}\right]\mathcal{F}_{n_e}(t,\frac{x}{t}).
\end{aligned}
\end{equation}
Utilizing the transformations described in eq. (\ref{eq:(3.10)}), the cross ratios in eq. (\ref{4.15b}) in the bipartite limit ($L\to \infty$) are given as follows
\begin{equation}
\lim_{L\to \infty} t= e^{-\frac{2\pi\xi}{\beta}}, \,\,\,
\lim_{L\to \infty} \frac{x}{t}=\frac{-2\pi}{\beta}\rho.
\end{equation}
Note that as described in \cite{Calabrese:2014yza} the replica limit must precede the bipartite limit in the evaluation of the four point twist correlator on the cylinder. Upon implementing the correct order of the
of the replica and the bipartite limits the four point twist correlator in eq. (\ref{4.20c}) may  finally be expressed as

\begin{equation}\label{4.26a}
\begin{aligned}
\mathcal{E}&=\frac{c_L}{4}\ln\left[\frac{\beta}{\pi a}\sinh\left(\frac{\pi\xi}{\beta}\right)\right ]+\frac{c_M}{4}\frac{\pi\rho}{\beta}\coth\left(\frac{\pi\xi}{\beta}\right)
-\frac{c_L}{4}\frac{\pi\xi}{\beta}-\frac{c_M}{4}\frac{\pi\rho}{\beta}\\
&+f(e^{-\frac{2\pi\xi}{\beta}},-\frac{2\pi\rho}{\beta})+2\ln k_{1/2}.
\end{aligned}
\end{equation}
Here $a$ is a UV cutof{}f and the arbitrary function $f(t,x/t)$ describing the non universal part 
is given as
\begin{equation}
f(t,x/t)\equiv\lim_{n_e\to 1} \ln[\mathcal{F}_{n_e}(t,x/t)]
\end{equation}
and the last term is a non universal constant. 
Note that using eq. (\ref{3.14a}), eq. (\ref{4.26a}) may be expressed in the following fashion
\begin{equation}
\begin{aligned}
\mathcal{E}&=\frac{3}{2} \big[ S_A-S_A^{th} \big] +f(e^{-\frac{2\pi\xi}{\beta}},-\frac{2\pi\rho}{\beta})+\mathrm{const.},
\end{aligned}
\end{equation}
where $S_A$ and $S_A^{th} $ denote the entanglement entropy and the thermal entropy corresponding to the 
mixed state described by the single interval in the $GCFT_{1+1}$ and the other terms are non universal and depend on the full operator content of the theory. Note as earlier, the elimination of the thermal entropy from the entanglement entropy in the universal part for the entanglement negativity conforms to its characterization as the distillable entanglement of the mixed state under consideration.
Remarkably we observe that the universal part of the entanglement negativity is identical in form to that of a relativistic $CFT_{1+1}$ given in eq. (\ref{2.19}). This illustrates the universality of the result and its sole dependence on the identical conformal structures of the two theories. This completes
our analysis for the entanglement negativity for the case of the zero and finite temperature bipartite pure and mixed state of a single interval in a $GCFT_{1+1}$.

\subsection{Entanglement negativity for adjacent intervals }

\begin{figure}[H]
\centering
\includegraphics[width=7cm]{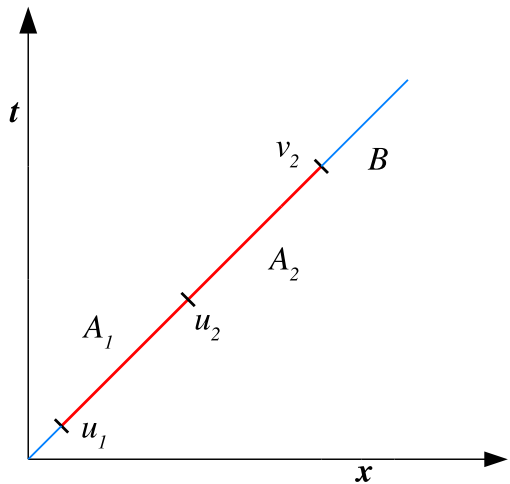}
\caption{Two adjacent intervals.}\label{fig4}
\end{figure}

Having described the various scenarios associated with a single interval in a $GCFT_{1+1}$ we now
focus on computing the entanglement negativity for the zero temperature mixed state configuration of two adjacent intervals in a $GCFT_{1+1}$. To this end we consider the adjacent limit $v_1\to u_2$ for the two disjoint intervals in figure \ref{fig3} to arrive at the configuration described in figure \ref{fig4}. For this configuration the right hand side of the eq. (\ref{eq:(5.1)}) reduces to a three point twist correlator on the GCFT plane as follows

\begin{equation}\label{eq:(5.6)}
\begin{aligned}
\textrm{Tr}(\rho_A^{T_2})^{n_e}=\left<\Phi_{n_e}(x_1,t_1)\Phi_{-n_e}^2(x_2,t_2)\Phi_{n_e}(x_3,t_3)\right>.
\end{aligned}
\end{equation}
Substituting the weights of the twist fields from
eq. (\ref{eq:(5.4)}) into eq. (\ref{eq:(5.6)}) we find
\begin{equation}
\begin{aligned}
\textrm{Tr}(\rho_A^{T_2})^{n_e}&=k_{n_e}(t_{12}t_{23})^{-(c_L/12)\left(\frac{n_e}{2}-\frac{2}{n_e}\right)}
t_{13}^{-(c_L/12)(1/n_e+n_e/2)}\\
&\times \exp\Bigg[-\frac{c_M}{12}\Big(\frac{n_e}{2}-\frac{2}{n_e}\Big) \left (\frac{x_{12}}{t_{12}}+\frac{x_{23}}{t_{23}} \right )
- \frac{c_M}{12}(1/n_e+n_e/2)\frac{x_{13}}{t_{13}}\Bigg].
\end{aligned}
\end{equation}
Utilizing eq. (\ref{eq: (5.1A}), the entanglement negativity for the mixed state of two adjacent intervals may be obtained as follows
\begin{equation}\label{4.28e}
\mathcal{E}=\frac{c_L}{8}\ln \left(\frac{t_{12}t_{23}}{a(t_{12}+t_{23})}\right)+\frac{c_M}{8}\left (\frac{x_{12}}{t_{12}}+\frac{x_{23}}{t_{23}}-\frac{x_{13}}{t_{13}}\right ) + \mathrm{const}.,
\end{equation}
where $a$ is a UV cutof{}f and the third term is a non universal constant for the three point function in the $GCFT_{1+1}$. 

We observe from eq. (\ref{eq:4.8}) that the universal part of the entanglement negativity in eq. (\ref{4.28e}) may be expressed as
\begin{equation}
\mathcal{E}=\frac{3}{4}\left(S_{A_1}+S_{A_2}-S_{A_1\cup A_2}\right)=\frac{3}{4}\,{\cal I}(A_1, A_2),
\end{equation}
where $ {\cal I}(A_1, A_2)$ is the universal part of the mutual information\footnote{The mutual information between the subsystem $A$ and $B$ of a bipartite system $A\cup B$ is defined as 
${\cal I}(A,B)=S_A+S_B-S_{A\cup B}$ where $S_A$ and $S_B$ are the entanglement entropy of subsystems $A$ and $B$ respectively. Note that the entanglement negativity and the mutual information are distinct measures in quantum information theory. For the configuration in question their universal parts are identical.} between the subsystems described by the intervals $A_1$ and $A_2$.

\subsection{Negativity for two adjacent intervals in vacuum for a finite size system}
It is now possible to extend the above analysis to compute the entanglement negativity for the zero temperature mixed state of adjacent intervals for a finite size system described by a $GCFT_{1+1}$ on a cylinder of circumference $L$. For this purpose it is required to evaluate the corresponding three point twist correlator on the cylinder of circumference $L$ utilizing the transformation of a primary field from the $GCFT$ plane to the cylinder described in eq. (\ref{3.18f}) to arrive at the following expression
\begin{equation}\label{4.31a}
\begin{aligned}
\left<\Phi_{n_e}(\xi_1,\rho_1)\Phi^2_{-n_e}(\xi_2,\rho_2)\Phi_{n_e}(\xi_3,\rho_3)\right>
&=\left(\frac{L}{2\pi i}\right)^{-2\Delta_{n_e}-\Delta_{n_e}^{(2)}} 
\exp\Bigg[\frac{2\pi i}{L}(\xi_1\Delta_{n_e}+\xi_2\Delta_{n_e}^{(2)}+\xi_3\Delta_{n_e}\\
&+\rho_1\chi_{n_e}+\rho_2\chi_{n_e}^{(2)}+\rho_3\chi_{n_e})\Bigg]\,
 t_{12}^{-\Delta_{n_e}^{(2)}}\,
    t_{23}^{-\Delta_{n_e}^{(2)}}\,
   t_{13}^{-(2\Delta_{n_e}-\Delta_{n_e}^{(2)})}\\
   &\exp\Bigg[- \chi_{n_e}^{(2)} \frac{x_{12}}{t_{12}} 
   - \chi_{n_e}^{(2)} \frac{x_{23}}{t_{23}}
   -(2\chi_{n_e}-\chi_{n_e}^{(2)})\frac{x_{13}}{t_{13}}\Bigg],
\end{aligned}
\end{equation}
where $(\Delta_{n_e},\chi_{n_e}), (\Delta_{n_e}^{(2)},\chi_{n_e}^{(2)})$ are the weights of the $GCFT$ twist fields $\Phi_{n_e}$ and $\Phi^2_{-n_e}$ respectively. Note that the right hand side of the above equation involves the three point twist correlator on the $GCFT$ plane apart from the leading exponential factor with its coefficient. We may now employ the coordinate transformations described in eq. (\ref{eq:(3.14)}) on the right hand side of the above eq. (\ref{4.31a}) and use eq. (\ref{eq: (5.1A}), to obtain the entanglement negativity of the zero temperature mixed state under consideration as follows
\begin{equation}\label{4.33}
\begin{aligned}
\mathcal{E}&=\frac{c_L}{8}\ln\left[\frac{L}{\pi a}\frac{\mathrm{sin}\left(\frac{\pi \xi_{12}}{L}\right)\mathrm{sin}\left(\frac{\pi \xi_{23}}{L}\right)}{\mathrm{sin}\left(\frac{\pi(\xi_{12}+\xi_{23})}{L}\right)}\right]
+\frac{c_M}{8}\frac{\pi}{L}\bigg [\rho_{12}\,\mathrm{cot}\Big(\frac{\pi\xi_{12}}{L}\Big)+\rho_{23}\,\mathrm{cot}\Big(\frac{\pi \xi_{23}}{L}\Big)\\
&-\rho_{13}\,\mathrm{cot}\Big (\frac{\pi\xi_{13}}{L}\Big ) \bigg ]+\mathrm{const}.,
\end{aligned}
\end{equation}
where $a$ is a UV cutof{}f. As earlier we observe from eq. (\ref{3.17A}) that the universal part of the entanglement negativity in eq. (\ref{4.33}) may be expressed in terms of the universal part of the mutual information ${\cal I}(A_1, A_2)$ between the subsystems $A_1$ and $A_2$ as follows
\begin{equation}
\mathcal{E}=\frac{3}{4}\left(S_{A_1}+S_{A_2}-S_{A_1\cup A_2}\right)= \frac{3}{4} {\cal I}(A_1, A_2).
\end{equation}

\subsection{Negativity for two adjacent intervals at a finite temperature}
In an exactly similar fashion as above we may now compute the entanglement negativity for the mixed state configuration of adjacent intervals in a $GCFT_{1+1}$ at a finite temperature $T$. To this end
it is required to evaluate the three point twist correlator in eq. (\ref{eq:(5.6)}) on a cylinder of circumference $\beta=1/T$. Utilizing the transformation for the $GCFT$ primary fields described in eq. (\ref{eq:(3.11)}) the three point twist correlator on the cylinder may be expressed as follows 
\begin{equation}\label{ft}
\begin{aligned}
\left<\Phi_{n_e}(\xi_1,\rho_1)\Phi^2_{-n_e}(\xi_2,\rho_2)\Phi_{n_e}(\xi_3,\rho_3)\right>_{\beta}
&=\left(\frac{\beta}{2\pi }\right)^{-2\Delta_{n_e}-\Delta_{n_e}^{(2)}} 
\exp\Bigg[\frac{2\pi}{\beta}(\xi_1\Delta_{n_e}+\xi_2\Delta_{n_e}^{(2)}+\xi_3\Delta_{n_e}\\
&+\rho_1\chi_{n_e}+\rho_2\chi_{n_e}^{(2)}+\rho_3\chi_{n_e})\Bigg]\,
 t_{12}^{-\Delta_{n_e}^{(2)}}\,
    t_{23}^{-\Delta_{n_e}^{(2)}}\,
   t_{13}^{-(2\Delta_{n_e}-\Delta_{n_e}^{(2)})}\\
   &\exp\Bigg[- \chi_{n_e}^{(2)} \frac{x_{12}}{t_{12}} 
   - \chi_{n_e}^{(2)} \frac{x_{23}}{t_{23}}
   -(2\chi_{n_e}-\chi_{n_e}^{(2)})\frac{x_{13}}{t_{13}}\Bigg],
\end{aligned}
\end{equation}
where $(\Delta_{n_e},\chi_{n_e}), (\Delta_{n_e}^{(2)},\chi_{n_e}^{(2)})$ are the weights of the $GCFT$ twist fields $\Phi_{n_e}$ and $\Phi^2_{-n_e}$ respectively. Note that as earlier the right hand side of the above equation involves the three point twist correlator on the $GCFT$ plane apart from the leading exponential factor with its coefficient. It is now possible to employ the coordinate transformation described in eq. (\ref{eq:(3.10)}) on the right hand side of the above eq. (\ref{ft}) and use eq. (\ref{eq: (5.1A}), to obtain the entanglement negativity of the finite temperature mixed state under consideration as follows
\begin{equation}\label{4.30}
\begin{aligned}
\mathcal{E}&=\frac{c_L}{8}\mathrm{ln}\left[\frac{\beta}{\pi a}\frac{\mathrm{sinh}\left(\frac{\pi \xi_{12}}{\beta}\right)\mathrm{sinh}\left(\frac{\pi \xi_{23}}{\beta}\right)}{\mathrm{sinh}\left(\frac{\pi(\xi_{12}+\xi_{23})}{\beta}\right)}\right]
+\frac{c_M}{8}\frac{\pi}{\beta}\bigg [\rho_{12}\,\mathrm{coth}\Big(\frac{\pi\xi_{12}}{\beta}\Big)+\rho_{23}\,\mathrm{coth}\Big(\frac{\pi \xi_{23}}{\beta}\Big)\\
&-\rho_{13}\,\mathrm{coth}\Big (\frac{\pi\xi_{13}}{\beta}\Big ) \bigg ]+\mathrm{const}.,
\end{aligned}
\end{equation}
where $a$ is a UV cutof{}f. As earlier we observe from eq. (\ref{3.14a}) that the universal part of the entanglement negativity in eq. (\ref{4.30}) may be expressed in terms of the mutual information ${\cal I}(A_1, A_2)$ as follows
\begin{equation}
\mathcal{E}=\frac{3}{4}\left(S_{A_1}+S_{A_2}-S_{A_1\cup A_2}\right)=\frac{3}{4}{\cal I}(A_1, A_2).
\end{equation}


\section{Summary and Conclusions}\label{sec5}

To summarize, in this article we have obtained the entanglement negativity for various bipartite pure and mixed state configurations in a class of $(1+1)$-dimensional Galilean conformal field theories. In this context utilizing a replica technique we have computed the entanglement negativity for the pure state configurations of a single interval in infinite and finite size systems described by a $GCFT_{1+1}$. It is observed that, as expected from quantum information theory the pure state entanglement negativity is given by the R\'{e}nyi entropy of order half. Interestingly our analysis reproduces the universal feature of pure state entanglement negativity in a relativistic $CFT_{1+1}$ where it is proportional to the corresponding entanglement entropy. This serves as a strong consistency check of our construction for characterizing the entanglement negativity of such bipartite pure states in a $GCFT_{1+1}$.

Subsequently we have investigated the interesting issue of determining the entanglement negativity for  mixed state configurations of adjacent intervals for zero, finite temperature and finite size in a $GCFT_{1+1}$. As earlier our results reproduce the universal feature of mixed state entanglement negativity to be proportional to the mutual information for this configuration in a relativistic $CFT_{1+1}$. Following this we have obtained a construction for the non trivial issue of the entanglement negativity for the finite temperature mixed state configuration of a single interval in an infinite system described by a $GCFT_{1+1}$. The replica technique for this scenario involved the determination of the partial transpose of the density matrix over an infinite cylinder which has been effected through a non trivial bipartite limit following the replica limit. Interestingly it was observed that the universal part of the entanglement negativity obtained through the above procedure was characterized by the elimination of the thermal contribution from the entanglement entropy for the finite temperature mixed state under consideration. As earlier the above result exactly reproduces the universal feature of entanglement negativity for the corresponding finite temperature mixed state in a relativistic $CFT_{1+1}$. Naturally, this also constitutes a strong consistency check of our construction for the entanglement negativity of both zero and finite temperature mixed states in a $GCFT_{1+1}$.

The systematic procedure to compute the entanglement negativity for mixed states prescribed by our construction should find interesting applications to entanglement issues in (1+1)-dimensional non relativistic systems with conformal symmetry in condensed matter physics. It would be extremely significant to characterize a holographic description of the entanglement negativity for $GCFT_{1+1}$ in the context of flat holography \cite{Bagchi:2010eg,Hijano:2017eii,Jiang:2017ecm,Fareghbal:2013ifa}. We hope to return to these exciting issues and applications in the near future.

\section{Acknowledgements}

We would like to thank Arjun Bagchi for useful references and for reading the manuscript.

\bibliographystyle{utphys}

\bibliography{gcftbib}

\end{document}